\documentclass{svproc}
\usepackage{url}

   \usepackage[update,prepend]{epstopdf} % automaticlaly converts .eps to .pdf file
   \usepackage{grffile} % allows graphics files to have "." in filename
    \usepackage{color}
    \usepackage{thumbpdf}    %
    \usepackage[ 
        final,
        breaklinks=true,
        colorlinks=true,           % coloured links 
        pdfborder={0 0 1},    % coloured box about link
        citecolor={blue},linkcolor={blue},urlcolor={red},
        citebordercolor={1 0 0},linkbordercolor={0 0 1},urlbordercolor={1 0 0},
        pdfpagemode=UseOutlines,% whether Acrobat opens with
        bookmarks=true,bookmarksopenlevel=4% write .pdf table of contents.
        ]{hyperref}         % Hyperlinks for PDF versions

   \usepackage{xparse}
   \NewDocumentCommand{\arxiv} %
   {r [: u{ [} u{]]} }{[\href{http://arxiv.org/abs/#2}{arXiv:#2}~[#3]]}
   \NewDocumentCommand{\arxivold} {r[]}{[\href{http://arxiv.org/abs/#1}{#1}]}
   \NewDocumentCommand{\arXiv} %
   {r [: u{ [} u{]]} }{[\href{http://arxiv.org/abs/#2}{arXiv:#2}~[#3]]}
   \NewDocumentCommand{\arXivold} {r[]}{[\href{http://arxiv.org/abs/#1}{#1}]}

   \usepackage{graphicx}          %%% for pictures and figures
\usepackage{multirow}                   %%% tabs over multiple rows
\usepackage{amssymb}
\usepackage{amsmath}
\usepackage{bbm} % double-line symbols (RR), with numerals (11) and lower-case
\usepackage{xspace}                    %%% correct spacing self-defined macros
\usepackage{dcolumn}                    %%% Align table columns on decimal point
\usepackage{bm}                        %%% bold math

\usepackage{cancel}	                  % produce backslash though symbol

\usepackage{rotating}

\usepackage{floatpag}           % eliminate page number from pages with
\floatpagestyle{plain}          % ordinary behaviour
\usepackage{slashed}
\newcommand{\blue}[1]{#1}
\newcommand{\red}[1]{#1}

\newcommand{\eg}{\textit{e.g.}\xspace}

\newcommand{\etal}{\textit{et al.}\xspace}

\newcommand{\hqq}{\hspace{1em}}

\newcommand{\half}{\frac{1}{2}}

\newcommand{\de}{\partial}

\newcommand{\MeV}{\ensuremath{\mathrm{MeV}}}

\newcommand{\ChiEFT}{$\chi$EFT\xspace}

\newcommand{\NXLO}[1]{N\ensuremath{{}^{#1}}LO\xspace}

\newcommand{\HIGS}{HI$\gamma$S\xspace}

\newcommand{\threeHe}{\ensuremath{{}^3}He\xspace}

\newcommand{\alphaep}{\ensuremath{\alpha_{E1}^{(\mathrm{p})}}}
\newcommand{\betamp}{\ensuremath{\beta_{M1}^{(\mathrm{p})}}}

\newcommand{\alphaen}{\ensuremath{\alpha_{E1}^{(\mathrm{n})}}}
\newcommand{\betamn}{\ensuremath{\beta_{M1}^{(\mathrm{n})}}}
\newcommand{\gammaeen}{\ensuremath{\gamma_{E1E1}^{(\mathrm{n})}}}
\newcommand{\gammammn}{\ensuremath{\gamma_{M1M1}^{(\mathrm{n})}}}
\newcommand{\gammaemn}{\ensuremath{\gamma_{E1M2}^{(\mathrm{n})}}}

\newcommand{\omegalab}{\ensuremath{\omega_\mathrm{lab}}}

\newcommand{\calO}{\mathcal{O}}

\begin{document}
\mainmatter              % start of a contribution
\title{Polarisabilities from Compton Scattering on \threeHe}
\titlerunning{Polarisabilities from Compton Scattering on \threeHe}  % abbreviated title (for running head)
\author{Harald W.~Grie\3hammer\inst{1} \and Judith A.~McGovern\inst{2}}
\authorrunning{Harald W.~Grie\3hammer and Judith A.~McGovern} % abbreviated author list (for running head)
\tocauthor{Harald W.~Grie\3hammer and Judith A.~McGovern}
\institute{%Institute for Nuclear Studies, 
  Department of Physics, The
    George Washington University, Washington DC% 20052
    , USA
\email{hgrie@gwu.edu}
\and School of Physics and Astronomy, The University of
    Manchester, Manchester% M13 9PL
, UK
}

\maketitle              % typeset the title of the contribution

\vspace*{-3.5ex}

\begin{abstract}
  This executive summary of recent theory progress in Compton scattering off
  \threeHe focuses on determining neutron polarisabilities; see
  ref.~\cite{Margaryan:2018opu} and references therein for details and a
  better bibliography\footnote{Prepared for the Proceedings of the \emph{22nd
    International Conference on Few-Body Problems in Physics}, Caen 9-13 July
  2018.}. 
% Contributed talks (15+5 mins):  4 pages
% We would like to encourage you to list your keywords within
% the abstract section using the \keywords{...} command.
\end{abstract}

\vspace*{-2ex}

%%%%%%%%%%%%%%%%%%%%%%%%%%%%%%%%%%%%%%%%%%%%%%%%%%%%%%%%%%%%%%%%%%%%%%%%%%%
%%%%%%%%%%%%%%%%%%%%%%%%%%%%%%%%%%%%%%%%%%%%%%%%%%%%%%%%%%%%%%%%%%%%%%%%%%%
%%%%%%%%%%%%%%%%%%%%%%%%%%%%%%%%%%%%%%%%%%%%%%%%%%%%%%%%%%%%%%%%%%%%%%%%%%%
% \section{}
% \label{sec:}
% %

\noindent
%
% J.~McGovern's plenary discusses the large-scale international effort, gains
% and goals of a new generation of high-precision facilities to 
% % understand low-energy Nuclear Physics by extracting 
% extract nucleon polarisabilities from Compton scattering experiments.
% M.~Elihau, R.~Higa, R.~Pohl and M.~S.~Safronova highlight the importance of
% electromagnetic polarisabilities in many contexts. C.~Howell shows
% that determining them by experiments takes years of planning, execution and
% analysis --- and commensurate theory support. 
Two plenaries discuss the large-scale international effort, gains
and goals of a new generation of high-precision facilities to 
% understand low-energy Nuclear Physics by extracting 
extract nucleon polarisabilities from Compton scattering experiments, and show
that determining them by experiments takes years of planning, execution and
analysis --- and commensurate theory support. Others highlight the importance
of electromagnetic polarisabilities in many contexts.
We thus refer to all these contributions~\cite{proc} for motivation and
context, and concentrate on theory progress for one target nucleus: \threeHe.

\paragraph{Setting the Stage}
Low-energy Compton scattering $\gamma X\to\gamma X$ probes a target's internal
degrees of freedom in the electric and magnetic fields of a real %, external
photon. These fields induce radiation multipoles by displacing the target
constituents. The angular and energy dependence of the emitted radiation
encodes information from the symmetries and strengths which govern the
interactions of the constituents with each other and with photons.
After subtracting the ``Born contributions'' (known from
one-photon data like form factors),
% one therefore parametrises the frequency-dependent stiffness of a nucleon $N$
% (spin $\frac{\vec{\sigma}}{2}$) against transitions of multipolarity
% $\blue{Xl\to Yl^\prime}$ ($l^\prime=l\pm\{0;1\}$; $X,Y=E,M$). When their rich
% information is compressed into the value at zero photon energy, these
% \emph{static polarisabilities}
its multipoles parametrise the stiffness of a nucleon $N$ (spin
$\frac{\vec{\sigma}}{2}$) against transitions % of multipolarity
$\blue{Xl\to Yl^\prime}$ at frequency $\omega$ ($l^\prime=l\pm\{0;1\}$;
$X,Y=E,M$; $T_{ij}=\half (\de_iT_j + \de_jT_i)$; $T=E,B$):
%\vspace*{-1ex}
%\begin{equation}
\[
\begin{split}
  \label{polsfromints}
  2\pi N^\dagger
  \big[&\red{\alpha_{E1}(\omega)}\vec{E}^2+
  \red{\beta_{M1}(\omega)}\vec{B}^2 +\red{\gamma_{E1E1}(\omega)}
  \vec{\sigma}\cdot(\vec{E}\times\dot{\vec{E}})
  +\red{\gamma_{M1M1}(\omega)}
  \vec{\sigma}\cdot(\vec{B}\times\dot{\vec{B}})
  \\&%
  %\non
  -2\red{\gamma_{M1E2}(\omega)}\sigma^iB^jE_{ij}+
  2\red{\gamma_{E1M2}(\omega)}\sigma^iE^jB_{ij} +(\mbox{higher multipoles})
  \big]N\;\;.
\end{split}
\]
%\end{equation} 
Six two-photon response functions suffice up to about $400\;\MeV$: two scalar
polarisabilities $\alpha_{E1}(\omega)$ and $\beta_{M1}(\omega)$ for
electric and magnetic dipole transitions; and
the four dipole spin-polarisabilities $\gamma_{E1E1}(\omega)$,
$\gamma_{M1M1}(\omega)$, $\gamma_{E1M2}(\omega)$,
$\gamma_{M1E2}(\omega)$. These test the nucleon-spin structure and complement
information from Jefferson Lab's spin programme. Intuitively, the
electromagnetic field of the spin degrees causes bi-refringence in the
nucleon, like in the classical Faraday-effect.
%
% Since they are related to the (real and virtual) excitation spectrum of the
% target, they probe the two-photon response of a nucleon, complementing the
% information available in the one-photon response (e.g.~in form
% factors). Differences between proton and neutron values stem from
% isospin-breaking interactions, exploring the interplay between chiral symmetry
% as well as the pattern of its breaking, and short-distance Physics.

The static values, $\alpha_{E1}\equiv\alpha_{E1}(\omega=0)$ etc., are often
just called \emph{``the'' polarisabilities} and condense the rich information
on the pion cloud, on the $\Delta(1232)$ excitation, and on the interplay
between chiral symmetry breaking and short-distance interactions.  These
fundamental quantities provide stringent tests for theoretical descriptions of
hadron structure. Moreover, they are ingredients to the neutron-proton mass
difference, the proton charge-radius puzzle, and the Lamb shift of muonic
hydrogen. To extract them, one must reliably extrapolate from data to
$\omega=0$.
Since pure neutron targets are unfeasible, nuclear binding and meson-exchange
effects must also be subtracted with reliable theory
uncertainties. Fortunately, Chiral Effective Field Theory (\ChiEFT) provides
model-independent estimates of higher-order corrections and encodes the
correct low-energy dynamics of QCD. For few-nucleon systems, it consistently
incorporates hadronic and nuclear currents, rescattering effects and wave
functions. The photon's interaction with the charged pion-exchange between
nucleons also probes few-nucleon binding. Even if scattering on a free neutron
were feasible, cross sections and signals for coherent scattering from nuclei
are markedly larger.

\paragraph{Elastic Compton scattering from \threeHe} is a promising means to
access neutron polarisabilities. In ref.~\cite{Choudhury:2007bh} and
subsequent publications, Shukla \etal showed that the differential cross
section between $50$ and $120\;\MeV$ is sensitive to
the electric and magnetic dipole polarisabilities of the neutron, 
$\alphaen$ and $\betamn$, and that scattering on polarised \threeHe provides
good sensitivity to the neutron spin polarisabilities.  This
% findings % were carried out at ${\cal O}(Q^3)$ in the Chiral Effective Field
% Theory expansion and
triggered several approved proposals at MAMI and \HIGS% to exploit this opportunity to extract
% neutron polarisabilities from elastic $\gamma\,$\threeHe scattering
.

%%%%%%%%%%%%%%%%%%%%%%%%%%%%%%%%%%%%%%%%%%%%%%%%%%%%%%%%%%%%%%%%%%%%%%%%%%%

\begin{figure}[!b]
 \includegraphics[clip=,height=0.34\linewidth]
{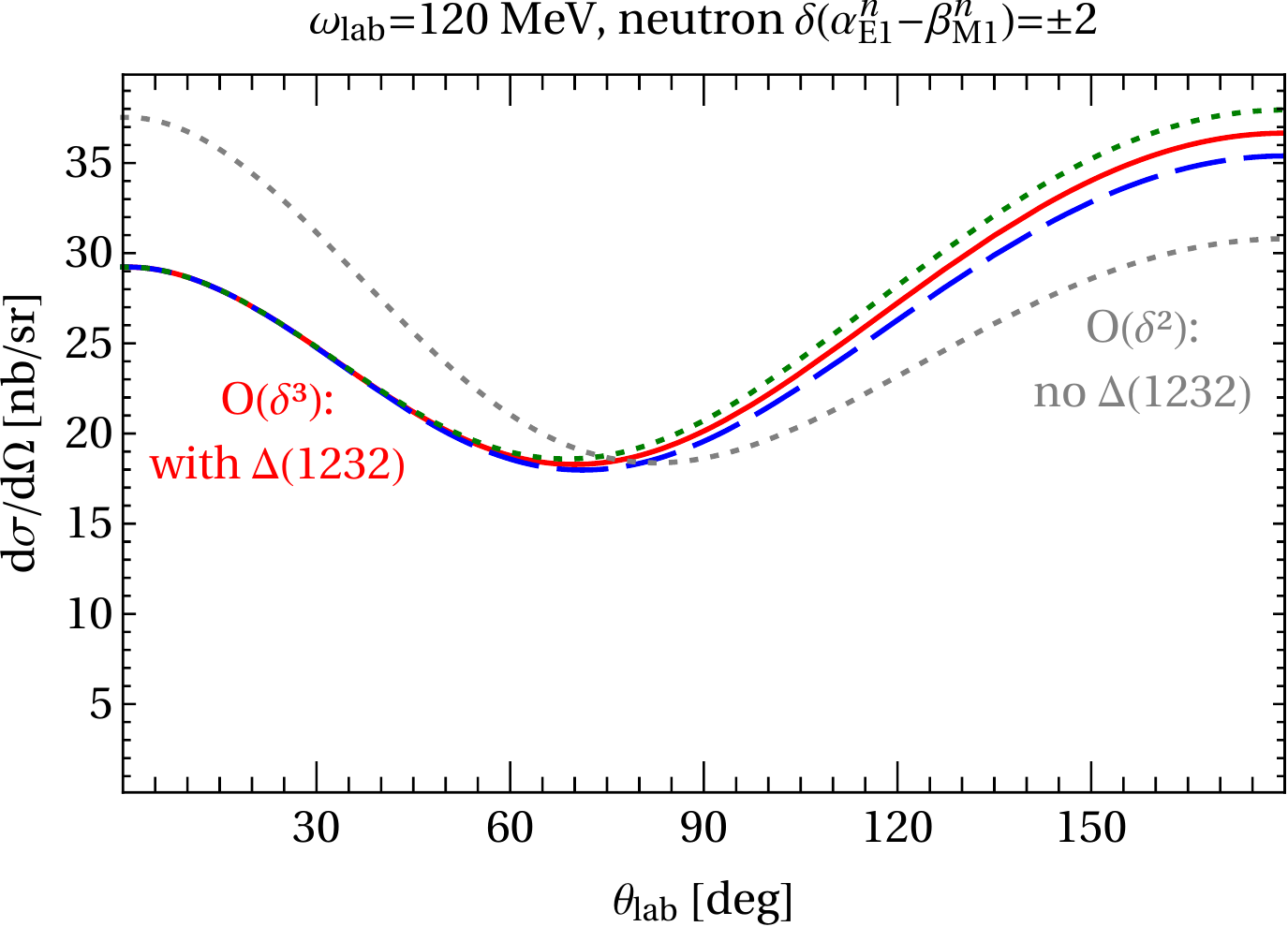}
\hqq
 \includegraphics[clip=,height=0.34\linewidth]
{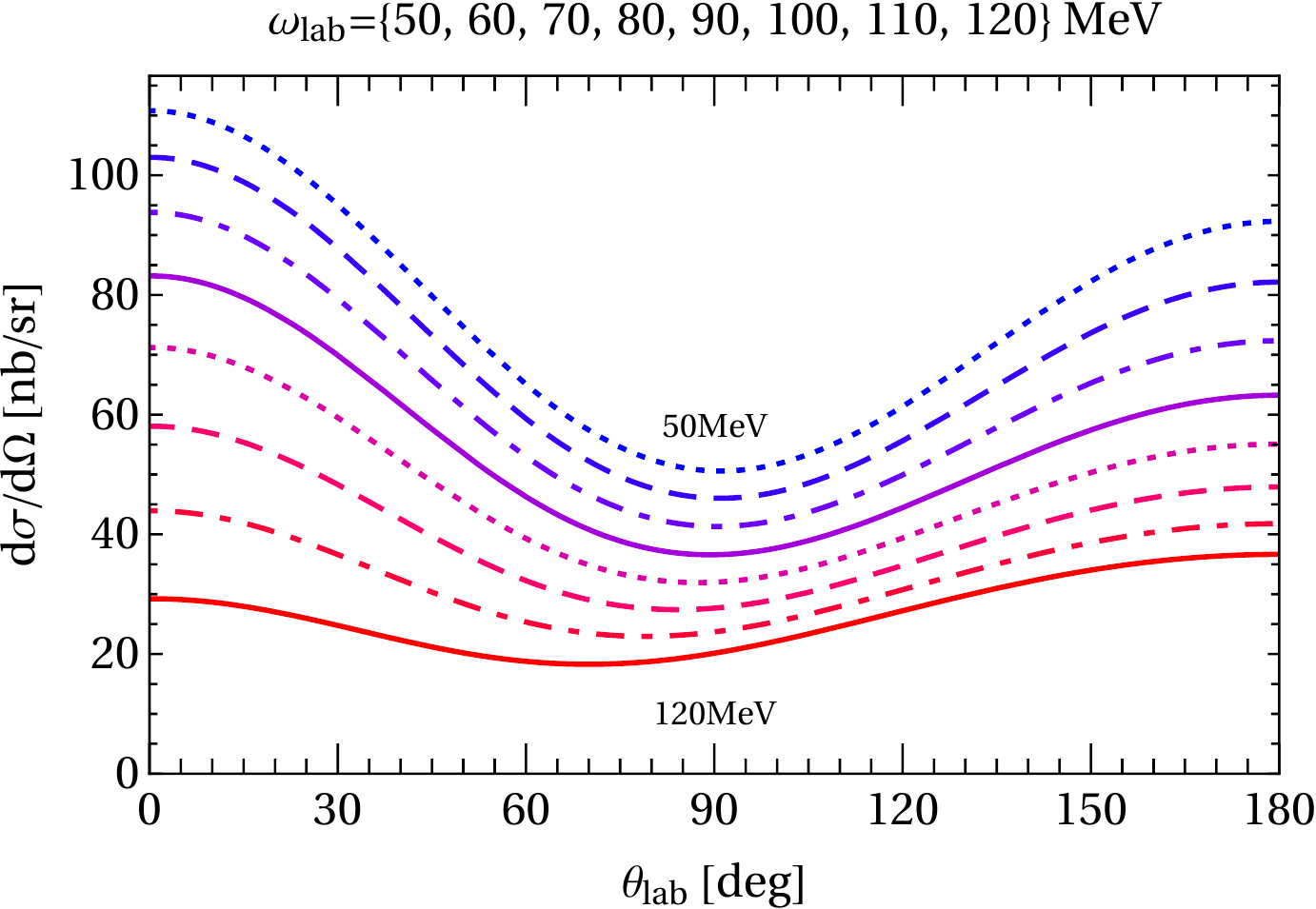}
\caption{\label{fig:1} Differential cross section.
% of \threeHe Compton scattering at $\omegalab=120\;\MeV$. 
  Left: at $\calO(e^2\delta^2)$ [no Delta] and $\calO(e^2\delta^3)$ [with
  Delta], and sensitivity to the neutron's scalar polarisabilities. Right:
  Energy dependence. }
\end{figure}

%%%%%%%%%%%%%%%%%%%%%%%%%%%%%%%%%%%%%

We recently extended these \ChiEFT predictions by one order to \NXLO{3}
[$\calO(e^2 \delta^3)$] by adding a dynamical Delta degree of freedom, and
provided results for photon lab energies between $50$ and $120\;\MeV$ for the
differential cross section, for the beam asymmetry $\Sigma_3$, and for the two
double asymmetries with circularly polarised photons and transversely or
longitudinally polarised targets, $%T_{11}^\mathrm{circ}=-\sqrt{2}\,
\Sigma_{2x}$ and $%T_{10}^\mathrm{circ}=
\Sigma_{2z}$.
These are the only non-zero observables below pion-production threshold in our
formulation.
We also found that the pioneering results were obtained from a computer code
which contained mistakes, triggering an erratum to
ref.~\cite{Choudhury:2007bh}.

% We employ the same $\gamma N$ and $\gamma NN$ amplitudes as in nucleon and
% deuteron Compton scattering.
At such energies, the complete photonuclear operator at \NXLO{3}
[${\cal O}(e^2 \delta^3)$] is: the Thomson and other
minimal-substitution terms; magnetic-moment couplings; dynamical
single-nucleon effects such as virtual pion loops and the Delta excitation;
and couplings of photons to the charged-pion exchange. All
terms are evaluated with \threeHe wave functions found from the same \ChiEFT
expansion.

\paragraph{Results} 
The dynamical Delta effects
% do not enter at low energies, but they
are obvious in all observables for $\omegalab\gtrsim100\;\MeV$; see
fig.~\ref{fig:1}. They markedly invert the fore-aft asymmetry of the cross
section and increase the magnitude of double asymmetries and their sensitivity
to spin polarisabilities, echoing similar findings for the deuteron.
The chiral expansion converges in this energy range quite well; see
\eg~fig.~\ref{fig:1}. The dependence on the choice of the \threeHe wave
function is small and can usually be distinguished from the effects of
polarisabilities by a different angular dependence.

\begin{figure}[!b]
 \includegraphics[clip=,height=0.38\linewidth,bb=0 0 345 251]
{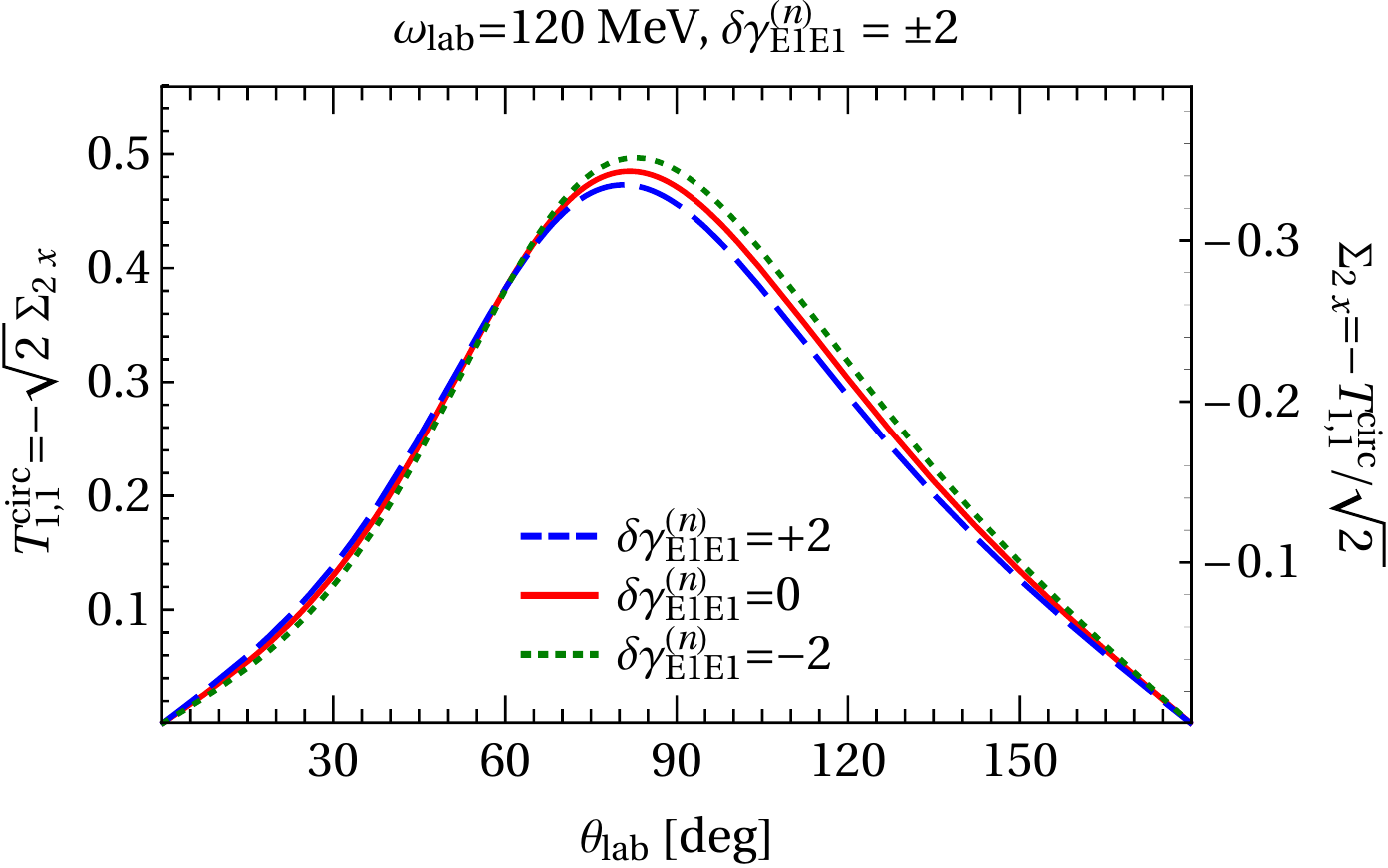}
%\hqq
 \includegraphics[clip=,height=0.38\linewidth,bb=44 0 345 251]
{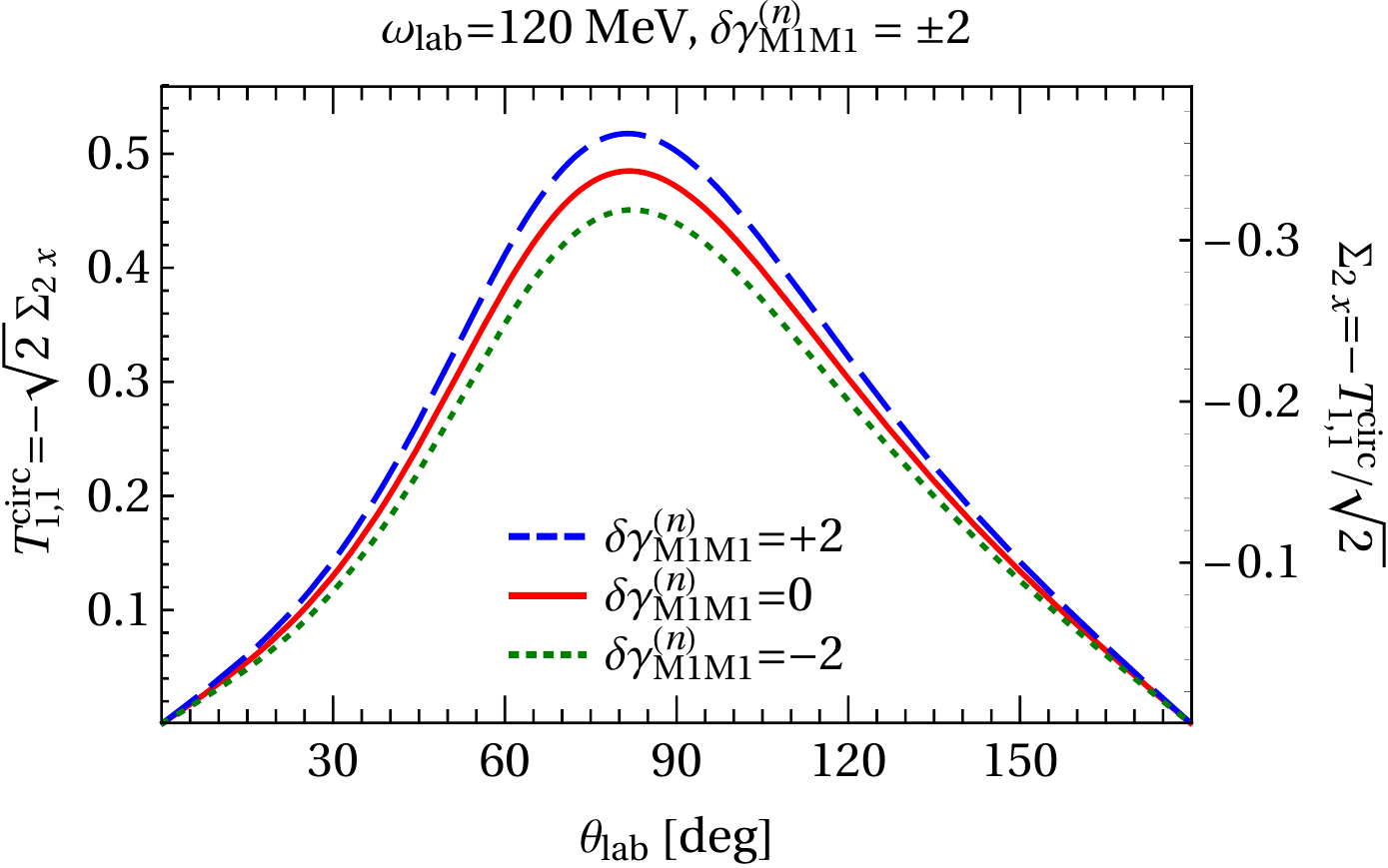}%
\caption{\label{fig:3} The sensitivity of $\Sigma_{2x}$ on the two 
  spin polarisabilities with the biggest impact. % at $\omegalab=120\;\MeV$.
}
\end{figure}

We found that $\alphaen - \betamn$ can be extracted from the cross section;
$%T_{11}^\mathrm{circ}=-\sqrt{2}\;
\Sigma_{2x}$
has a non-degenerate sensitivity to $\gammammn$ around $90^\circ$; and
$%T_{10}^\mathrm{circ}=
\Sigma_{2z}$ to $\gammaeen$ and $\gammaemn$; see fig.~\ref{fig:3}.
The beam asymmetry $%\Sigma^\mathrm{lin}=
\Sigma_3$
is dominated by the single-nucleon Thomson term and not very useful to
directly determine polarisabilities.
Ultimately, the most accurate polarisabilities will be inferred from data of
all four observables. For the spin polarisabilities, data at 
$\omegalab\gtrsim100\;\MeV$ will be crucial.

This exploration is part of an ongoing dialogue with our experimental
colleagues on the best kinematics and observables to extract neutron
polarisabilities. An interactive \emph{Mathematica} notebook is available from
\texttt{hgrie@gwu.edu}.
%Results are available as interactive \emph{Mathematica}
%notebook from \texttt{hgrie@gwu.edu}, % , including cross
% sections, rates and asymmetries when the scalar and spin polarisabilities
% are varied (including sum rule constraints).
%and quite robust. 
Results are quite robust.
Varying the single-nucleon amplitudes of complementary approaches like
dispersion relations will lead to sensitivities which are hardly discernible
from ours. Once data exist, a polarisability extraction will of course
need to address residual % theory
uncertainties with more diligence; see \eg~ref.~\cite{Griesshammer:2015ahu}.
%%%%%%%%%%%%%%%%%%
%%%%%%%%%%%%%%%%%%%%%%

%%%%%%%%%%%%%%%%%%

\begin{figure}[!htb]
 \includegraphics[clip=,height=0.34\linewidth,bb=0 0 378 263]
{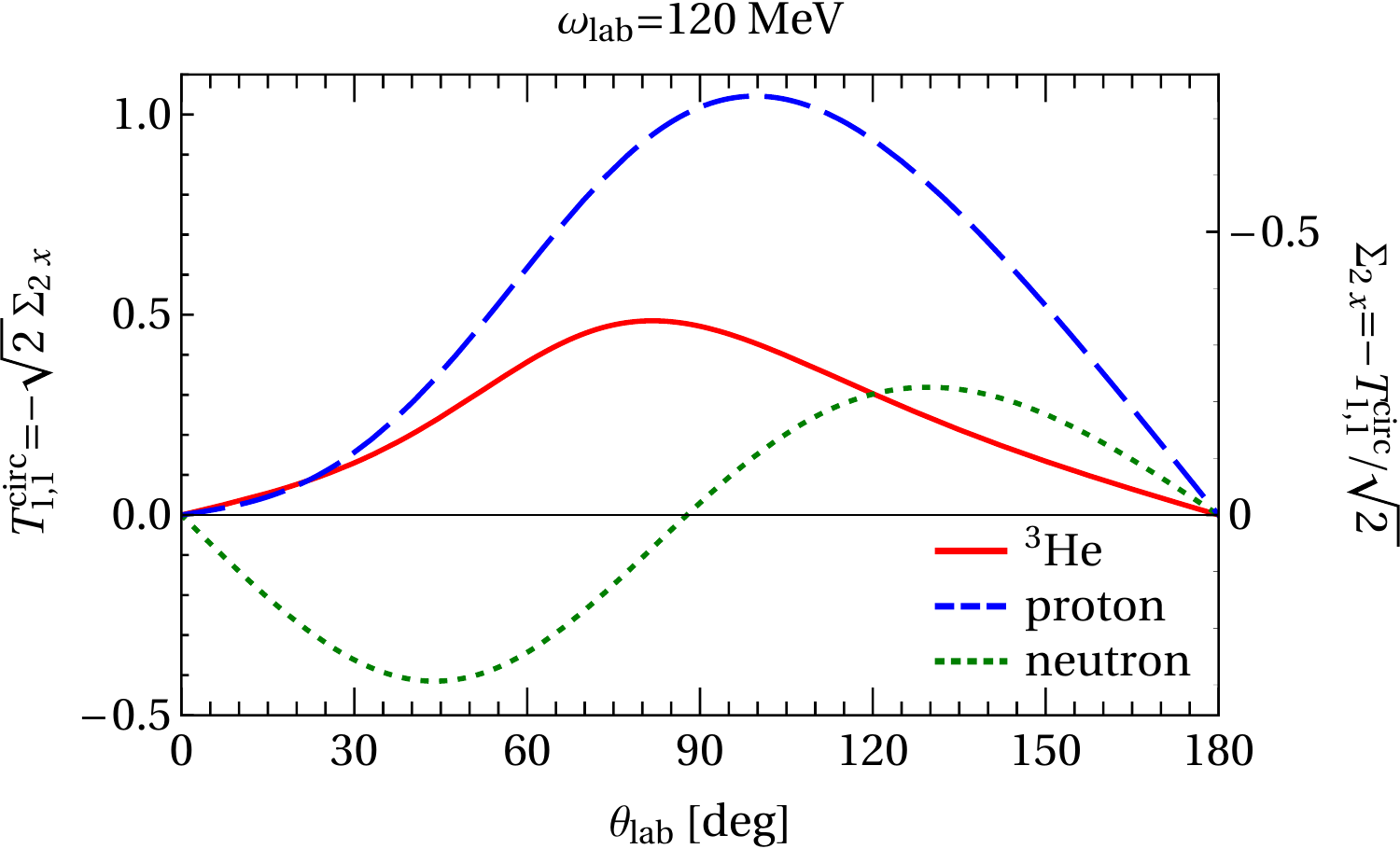}%
\hqq%
\includegraphics[clip=,height=0.34\linewidth,bb=0 0 390 273]
{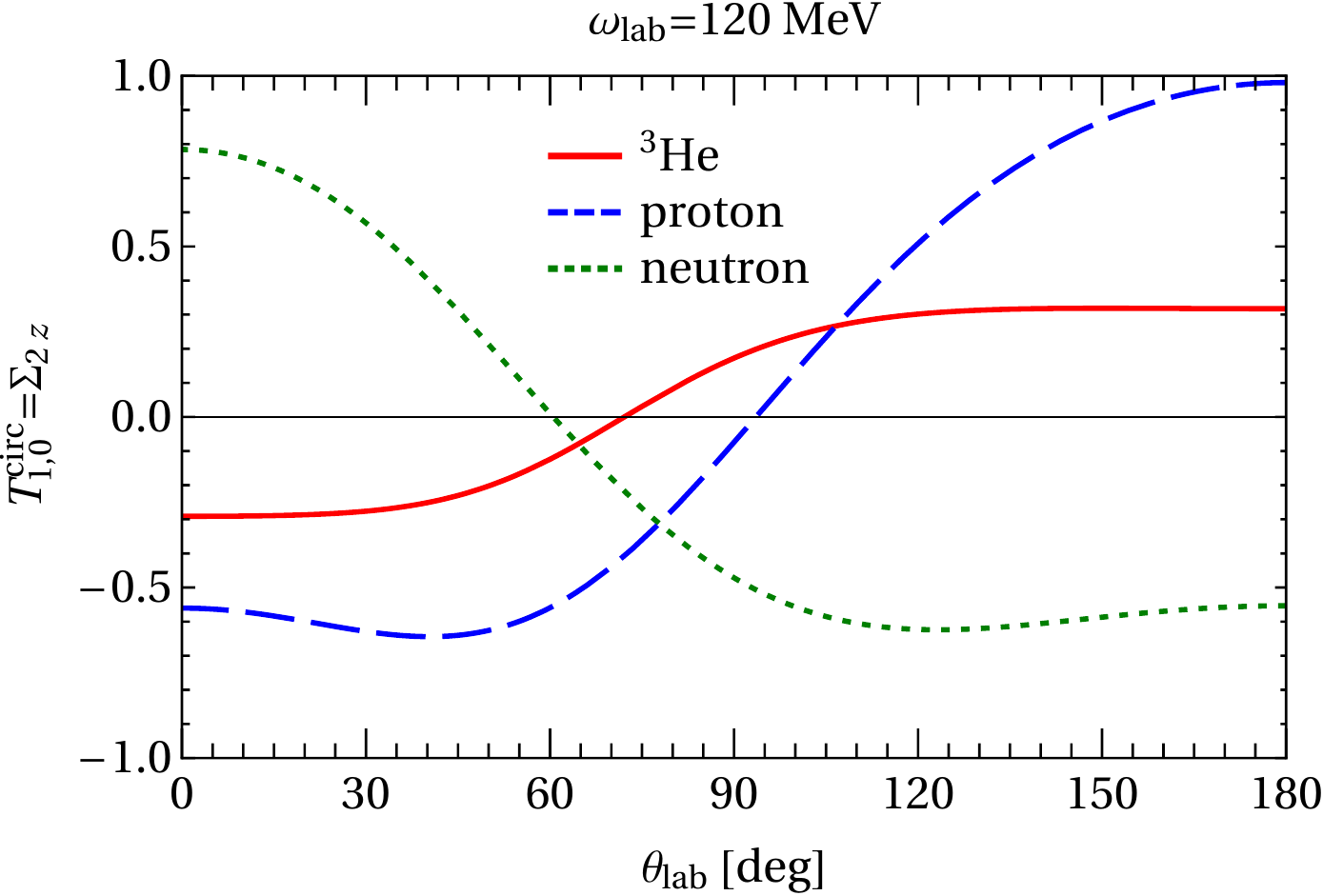}
\caption{\label{fig:2} The double asymmetries with circularly polarised beam
  and transversely (left) or longitudinally (right) polarised target, for the
  proton, neutron and \threeHe.}
\end{figure}

\paragraph{Nuclear Binding}
\ChiEFT also quantifies the angle- and energy-dependent corrective to the
na\"ive \threeHe picture as the sum of two protons with antiparallel spins and
one neutron. Sensitivity to the scalar polarisabilities enters indeed
approximately via $2 \alphaep + \alphaen$ and $2 \betamp + \betamn$, and the
double-asymmetries are $10$-to-$20$ times more sensitive to the spin
polarisabilities of the neutron than of the proton. However, fig.~\ref{fig:2}
confirms that there is no energy where polarised \threeHe simply acts as a
``free neutron-spin target''. The \emph{sensitivities}
% of
% $%T_{10}^{\rm circ}
% \Sigma_{2x}$ and $%T_{11}^{\rm circ}
% \Sigma_{2z}$
to neutron spin polarisabilities closely mimic those of free-neutron
observables. But their \emph{magnitudes} do not.

% Treating the process in
An impulse approximation would thus omit a key mechanism:
charged pion-ex\-change currents. % Their interference with the targeted
% neutron-structure effects is large, and neglecting them severely distorts
% extractions of nucleon polarisabilities.
Without their large interference with the % neutron 
polarisabilities, results
are severely distorted.
%%%%%%%%%%%%%%%%%%
The \ChiEFT expansion provides quantitative predictions % of this and
of the two-body currents, with reliable theory uncertainties. Detailed checks
of the convergence of the expansion for exchange currents and for the other
pieces of the \threeHe-Compton amplitude by performing a \NXLO{4}
[$\calO(e^2 \delta^4)$] calculation and extending the applicable energy range
are under way. They will allow for even more accurate extractions of
polarisabilities from upcoming data.

%%%%%%%%%%%%%%%%%%%%%%%%%%%%%%%%%%%%%%%%%%%%%%%%%%%%%%%%%%%%%%%%%%%%%%%%%%%%%%%
%%%%%%%%%%%%%%%%%%%%%%%%%%%%%%%%%%%%%%%%%%%%%%%%%%%%%%%%%%%%%%%%%%%%%%%%%%%%%%%
\paragraph{Acknowledgements}
%
%I am particularly indebted to J.~A.~McGovern 
HWG is particularly indebted to JMcG 
for filling in at 
% the conference
the oral presentation 
on very short notice. We are also grateful to
% This work reflects the collaboration
% with % her and
% D.~R.~Phillips and 
the other co-authors of ref.~\cite{Margaryan:2018opu}. 
% All insight is theirs, all mistakes mine.
This work was supported in part by the US Department of Energy under contract
DE-SC0015393 (HWG), and by UK Science and Technology Facilities Council grants
ST/L005794/1 and ST/P004423/1 (JMcG).

%%%%%%%%%%%%%%%%%%%%%%%%%%%%%%%%%%%%%%%%%%%%%%%%%%%%%%%%%%%%%%%%%%%%%%%%%%%%%%%
%%%%%%%%%%%%%%%%%%%%%%%%%%%%%%%%%%%%%%%%%%%%%%%%%%%%%%%%%%%%%%%%%%%%%%%%%%%%%%%
%%%%%%%%%%%%%%%%%%%%%%%%%%%%%%%%%%%%%%%%%%%%%%%%%%%%%%%%%%%%%%%%%%%%%%%%%%%%%%%

\end{document}